\documentclass[%
 reprint,
superscriptaddress,
 amsmath,amssymb,
 aps,prl,
floatfix,
]{revtex4-2}

\usepackage{graphicx}
\usepackage{dcolumn}
\usepackage{bm}
\usepackage{braket}
\usepackage{hyperref}
\hypersetup{pdfpagemode=FullScreen,
colorlinks=true, citecolor = blue, linkcolor=blue, filecolor=blue}

\usepackage{amsmath}
\usepackage[T1]{fontenc}

\usepackage{color}

\usepackage[normalem]{ulem}

\begin{document}

\preprint{APS/123-QED}

\title{Ultra-sensitive solid-state organic molecular microwave quantum receiver}

\author{Bo Zhang}
\email{bozhang_quantum@bit.edu.cn}
\affiliation{%
These authors contributed equally to this work
}%

\affiliation{%
Center for Quantum Technology Research and Key Laboratory of Advanced Optoelectronic Quantum Architecture and Measurements (MOE), Beijing Institute of Technology, Beijing 100081, China
}%
\affiliation{%
School of Physics, Beijing Institute of Technology, Beijing 100081, China
}%

\author{Yuchen Han}%
\affiliation{%
These authors contributed equally to this work
}%
\affiliation{%
Center for Quantum Technology Research and Key Laboratory of Advanced Optoelectronic Quantum Architecture and Measurements (MOE), Beijing Institute of Technology, Beijing 100081, China
}%
\affiliation{%
School of Physics, Beijing Institute of Technology, Beijing 100081, China
}%

\author{Hong-Liang Wu}%
\affiliation{%
These authors contributed equally to this work
}%
\affiliation{%
Center for Quantum Technology Research and Key Laboratory of Advanced Optoelectronic Quantum Architecture and Measurements (MOE), Beijing Institute of Technology, Beijing 100081, China
}%
\affiliation{%
School of Integrated Circuits and Electronics, Beijing Institute of Technology, Beijing 100081, China
}%

\author{Hao Wu}%
\email{hao.wu@bit.edu.cn}
\affiliation{%
Center for Quantum Technology Research and Key Laboratory of Advanced Optoelectronic Quantum Architecture and Measurements (MOE), Beijing Institute of Technology, Beijing 100081, China
}%
\affiliation{%
School of Physics, Beijing Institute of Technology, Beijing 100081, China
}%

\author{Shuo Yang}%
\affiliation{%
Center for Quantum Technology Research and Key Laboratory of Advanced Optoelectronic Quantum Architecture and Measurements (MOE), Beijing Institute of Technology, Beijing 100081, China
}%
\affiliation{%
School of Physics, Beijing Institute of Technology, Beijing 100081, China
}%

\author{Mark Oxborrow}
\affiliation{%
Department of Materials, Imperial College London, South Kensington SW7 2AZ, London, United Kingdom}%

\author{Qing Zhao}
\affiliation{%
Center for Quantum Technology Research and Key Laboratory of Advanced Optoelectronic Quantum Architecture and Measurements (MOE), Beijing Institute of Technology, Beijing 100081, China
}%
\affiliation{%
School of Physics, Beijing Institute of Technology, Beijing 100081, China
}%
\affiliation{%
Beijing Academy of Quantum Information Sciences, Beijing 100193, China
}%

\author{Yue Fu}
\affiliation{%
School of Integrated Circuits and Electronics, Beijing Institute of Technology, Beijing 100081, China
}%
\affiliation{%
MIIT Key Laboratory for Low-Dimensional Quantum Structure and Devices, Beijing Institute of Technology, Beijing 100081, China
}%

\author{Weibin Li}
\affiliation{%
School of Physics and Astronomy, and Centre for the Mathematics and Theoretical Physics of Quantum Non-equilibrium Systems, The University of Nottingham, Nottingham NG7 2RD, United Kingdom
}%

\author{Yeliang Wang}
\affiliation{%
School of Integrated Circuits and Electronics, Beijing Institute of Technology, Beijing 100081, China
}%
\affiliation{%
MIIT Key Laboratory for Low-Dimensional Quantum Structure and Devices, Beijing Institute of Technology, Beijing 100081, China
}%

\author{Dezhi Zheng}
\email{zhengdezhi@bit.edu.cn}
\affiliation{%
MIIT Key Laboratory of Complex-field Intelligent Exploration, Beijing Institute of Technology, Beijing, China
}%
\affiliation{%
Yangtze Delta Region Academy of Beijing Institute of Technology (Jiaxing), Jiaxing 314019, China
}%

\author{Jun Zhang}
\affiliation{%
MIIT Key Laboratory of Complex-field Intelligent Exploration, Beijing Institute of Technology, Beijing, China
}%
 
\begin{abstract}
High-accuracy microwave sensing is widely demanded in various fields, ranging from cosmology to microwave quantum technology. Quantum receivers based on inorganic solid-state spin systems are promising candidates for such purpose because of the stability and compatibility, but their best sensitivity is currently limited to a few pT/$\sqrt{\rm{Hz}}$. Here, by utilising an enhanced readout scheme with the state-of-the-art solid-state maser technology, we develop a robust microwave quantum receiver functioned by organic molecular spins at ambient conditions. Owing to the maser amplification, the sensitivity of the receiver achieves 6.14 ± 0.17 fT/$\sqrt{\rm{Hz}}$ which exceeds three orders of magnitude than that of the inorganic solid-state quantum receivers. The heterodyne detection without additional local oscillators improves bandwidth of the receiver and allows frequency detection. The scheme can be extended to other solid-state spin systems without complicated control pulses and thus enables practical applications such as electron spin resonance spectroscopy, dark matter searches, and astronomical observations.
\end{abstract}

\maketitle

\section{I. introduction}
Quantum sensing utilizes highly coherent and precisely controlled quantum systems to detect weak signals with unparalleled sensitivity and accuracy \cite{degen2017quantum}. Most recently, semiconductor defect color centers \cite{1974MolPh..27.1521A,2016PNAS..11314133B,2020Sci...370.1309B,2013NanoL..13.2073C} provide increasingly favorable platforms for ultra-sensitive detection of microwave (MW) magnetic fields \cite{Wang:2022klu,Meinel2021,Joas2017,Stark2017} with the advantages over other systems in terms of the robustness \cite{Liu2019}, biocompatibility \cite{Shi2018} and spatial resolution \cite{Taylor2008}. A typical approach for the improvement of the sensitivity is to increase the achievable number $N$ of sensing spins, because the projection noise is inversely proportional to $\sqrt{N}$ \cite{Taylor2008}. However, at high spin densities, other paramagnetic impurities become a sizable source of spin dephasing, such as P1 centers in diamond \cite{doi:10.1126/science.1155400,PhysRevB.64.041201}, severely limiting the sensitivity of the sensors. For MW field sensing, the uniform requirement of sophisticated control pulses also limits the available $N$ \cite{Stark2017}.

Here, we present a novel approach to ultra-sensitive MW magnetic-field detection, exploiting molecular electron spins in an organic crystal as quantum sensors. We focus on spins associated with photoexcited triplets of pentacene molecules doped in an organic single crystal when coupled to an electromagnetic mode of a microwave resonator (Fig. \ref{FIG.1}a), because they can be optically polarized and detected with maser technique, and exhibit excellent coherent properties even at room temperature \cite{2024arXiv240207572M,2024arXiv240213898S}. This hybrid quantum device can be used for room-temperature maser action \cite{Oxborrow2012}, strong spin-photon coupling  \cite{breeze2017room} and dc-magnetometry \cite{doi:10.1126/sciadv.ade1613}. In what follows, we show such a system can also be used for the precision sensing of microwave magnetic fields without complicated control pulses.

We discuss two merits of such hybrid quantum sensors. First, by exploiting the stimulated emission (i.e. the masing process) of the pentacene triplet spins which resonantly interact with the incoming microwave magnetic fields, the detected microwave signals undergo quantum amplification maximum to 17 dB so that we boost the sensitivity of the receiver to 6.14 ± 0.17 fT/$\sqrt{\rm{Hz}}$, which significantly outperforms the cutting-edge solid-state microwave quantum magnetometers based on NV centers  \cite{Wang:2022klu,PhysRevApplied.19.054095}. Second, the cQED effect of masing process enables the `classical-local-oscillator-free' heterodyne detection of the microwave signals far detuned from the cavity bandwidth with a sensitivity even better than that obtained under the resonant condition (Fig. \ref{FIG.1}a). Additionally, with the absence of paramagnetic impurities, high-density organic molecular crystals can be grown large ($cm-$scale) at low cost, where the sensitivity of molecular magnetometer can be substantially improved with scaling of the number of molecular sensors $1/\sqrt{N}$.

\section{II. results}
\subsection*{A. Molecular spin-based hybrid quantum system}

\begin{figure}[htbp!]
\centering
\includegraphics[width=8.8cm]{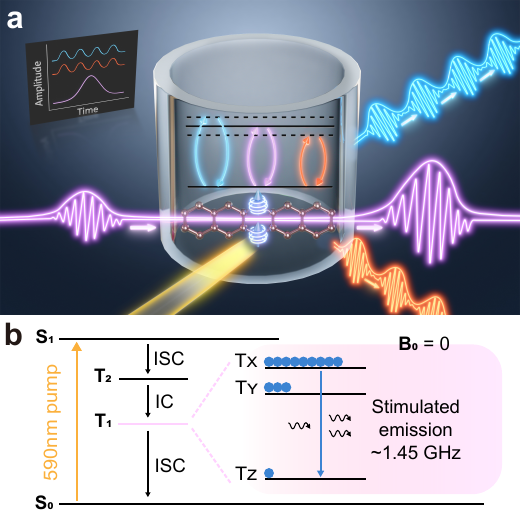}
\caption{\label{FIG.1}
\textbf{The principle of a microwave quantum receiver.} (\textbf{a}) Schematic depiction of the microwave quantum receiver. A pentacene-doped \textit{p}-terphenyl crystal is housed within a hollow cylinder of STO. Depending on the frequency of the incident microwaves, stimulated emission occurs under resonant (purple) or off-resonant (blue and orange) conditions, generating amplified or amplitude-modulated output signals. The inset demonstrates the typical outputs of the receiver. See text for details. (\textbf{b}) Jablonski diagram of the spin energy level structure of the pentacene molecules doped in \textit{p}-terphenyl. Upon excitation, highly spin-polarized states are formed between the triplet sublevels $\rm{T}_X$ and $\rm{T}_Z$. The input of 1.45-GHz microwave signals can trigger the stimulated emission, resulting in microwave amplification.
 }
\end{figure}

The microwave quantum receiver is functioned by a solid-state hybrid quantum system, in which an optically addressing molecular spin ensemble is coupled to the $\rm{TE}_{01\delta}$ mode of a three-dimensional dielectric microwave cavity. The spin ensemble is consisted of the triplet spins of pentacene molecules doped in the crystalline matrix of \textit{p}-terphenyl. 

As shown in Fig. \ref{FIG.1}b, under optical pumping at a wavelength of 590 nm, the electronic spins of the pentacene molecules are promoted from the singlet ground state $\rm{S}_0$ to the singlet excited state $\rm{S}_1$, and then undergo the transitions to the triplet state $\rm{T}_1$ with spin selectivity through the intersystem crossing (ISC) and internal conversion (IC), successively. At zero field, the whole process results in a non-Boltzmann distribution of populations in the non-degenerate sublevels of $\rm{T}_1$ at room temperature, where the population ratio of $\rm{T}_X$:$\rm{T}_Y$:$\rm{T}_Z$ equals to 0.76:0.16:0.08 \cite{sloop1981electron}. $\rm{T}_X$ and $\rm{T}_Z$ thus constitute a inverted two-level system with a transition frequency around 1.45 GHz. By resonantly coupling the pentacene molecules to the $\rm{TE}_{01\delta}$ mode of a high-Purcell-factor microwave cavity which is made of crystalline strontium titanate (STO) \cite{breeze2015enhanced}, a molecular spin-based hybrid quantum system is established and capable of highly sensitive microwave detection via measuring the magnetic field strengths of incident microwaves. The high sensitivity is achieved due to the intriguing feature of the hybrid quantum system that the received microwave photons are not only efficiently \textit{stored} in the cavity with a high quality factor $Q$, but also substantially \textit{amplified} via the stimulated emission of radiation, i.e. the masing process, between $\rm{T}_X$ and $\rm{T}_Z$. 

\subsection*{B. Calibration of microwave magnetic field strengths}

\begin{figure}[htbp!]
\centering
\includegraphics[width=8.8cm]{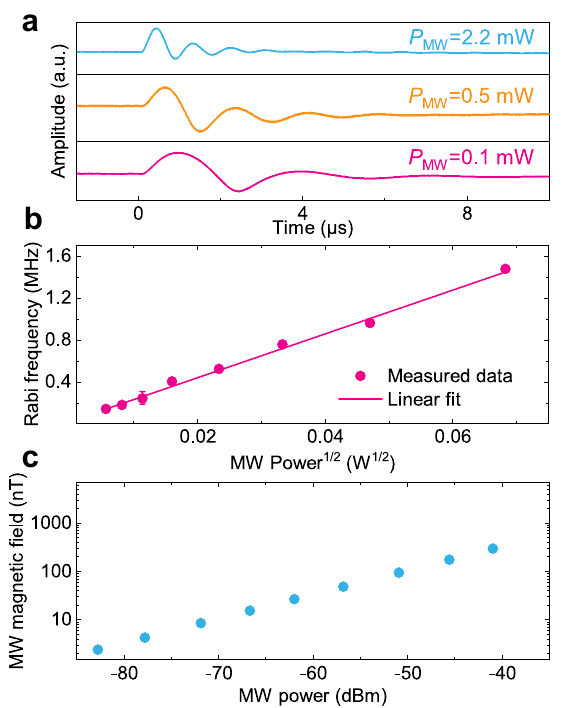}
\caption{\label{FIG.2}
\textbf{Magnetic field calibration.} (\textbf{a}) Rabi oscillation at different power levels. (\textbf{b}) The obtained Rabi frequencies show a linear relationship with the square roots of the power. A slope of 20.86 MHz/$\sqrt{\rm{W}}$ and a microwave magnetic-field conversion factor \textit{C} of 1.05 mT/$\sqrt{\rm{W}}$ can be derived. Each data point represents the mean ± standard deviation obtained by repeating three individual experiments. (\textbf{c}) The obtained conversion factor \textit{C} can be used to determine the microwave magnetic-field strength at different power levels. Error bars are contained within the data points.}
\end{figure}

As described above, the proposed microwave quantum receiver performs as a microwave magnetometer, of which the working principle is based on the detection of microwave magnetic fields. In order to evaluate the magnetometer's sensitivity, we first calibrated test magnetic fields of known microwave powers. The experimental setup is exhibited in supplementary materials. The test fields were produced by a microwave source whose output was controlled by a microwave switch synchronized to a laser with TTL pulses from an arbitrary signal generator (ASG). The  strength of the test magnetic field $B_1$ can be varied by adjusting the microwave power inputting the cavity $P_{\rm{MW}}$ with the microwave source and the variable attenuator in the circuit. The correlation between the microwave power and the magnetic field can be quantified by a conversion factor $C = B_1/\sqrt{P_{\rm{MW}}}$ \cite{poole1983electron}, which is necessary for calibrating the test fields.

The conversion factor is theoretically determined as $C = \sqrt{\frac{2\mu_0Q_{\rm L}}{V_{\rm{m}}\omega_0}}$ \cite{poole1983electron}, where $\mu_0 = 4\pi\times 10^{-7}$ H/${\rm{m}}$ is the vacuum permeability, $Q_{\rm{L}}$, $V_{\rm{m}}$ and $\omega_0$ are the loaded quality factor, magnetic mode volume and resonant frequency of a microwave cavity, respectively. The definition of the conversion factor indicates that $C$ can be predicted by characterizing the microwave cavity employed in our receiver. The loaded quality factor of the cavity resonant at $\omega_{\rm{c}} = 2\pi \times 1.4494$ GHz was measured to be $Q_{\rm{L}} = 1706$ via the transmission method. The magnetic mode volume $V_{\rm{m}}\approx5.4\times10^{-7}$ $\rm{m}^3$ was determined by means of the two-dimensional (2D)-axisymmetric finite-element simulation \cite{oxborrow2007traceable}. Thus, the conversion factor $C$ can be calculated to be 0.94 mT/$\sqrt{\rm{W}}$.

By measuring the transient electron paramagnetic resonance (trEPR) on the pentacene triplet spin ensemble, the conversion factor is experimentally measured according to the frequency of the $B_1$-induced Rabi oscillation $\Omega_{\rm{Rabi}}$ \cite{weber2018transient}, as shown in Fig. \ref{FIG.2}a. Based on $C = B_1/\sqrt{P_{\rm{MW}}}$ and $\Omega_{\rm{Rabi}} = -\gamma_{\rm{e}} B_1/\sqrt{2}$ \cite{Wang:2022klu}, where $\gamma_{\rm{e}} = -28$ MHz/mT is the gyromagnetic ratio of the pentacene triplet electron spin, the Rabi frequencies $\Omega_{\rm{Rabi}} = -\gamma_{\rm{e}} C \sqrt{P_{\rm{MW}}}/\sqrt{2}$ reveal a linear correlation with the square root of the microwave power inputting the cavity, and the conversion factor can be determined from the slope. By varying the microwave power set in the trEPR experiment, the corresponding changes of the observed Rabi frequency are demonstrated in Fig. \ref{FIG.2}b. The expected linear correlation is verified and well fit with the conversion factor $C=$ 1.05 ± 0.03 mT/$\sqrt{\rm{W}}$ which agrees well with the calculated value based on the parameters of the cavity.

\begin{figure*}[htbp!]
\centering
\includegraphics[width=1\textwidth]{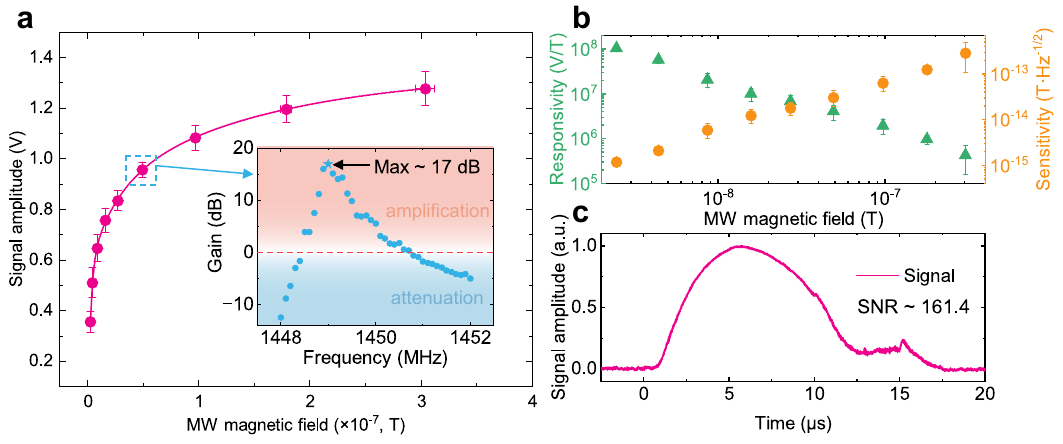}
\caption{\label{FIG.3}
\textbf{The sensitivity of microwave quantum receiver.} (\textbf{a}) The data points represent the signal strengths measured at different microwave magnetic fields and are fitted by an exponential equation. Inset: The maser amplification gain obtained at different frequencies with an input microwave magnetic-field strength of 49.23 nT. (\textbf{b}) The receiver's responsivities (blue triangles) at different magnetic fields determined from the slopes of the fitting curve in (a). The corresponding sensitivities at different magnetic fields are represented by purple circles. The error bars represent the standard errors of the data points in \textbf{a} and \textbf{b}. (\textbf{c}) A signal waveform with a SNR of 161.4 obtained at the lowest input microwave magnetic field corresponding to the lowest data point in (a).}
\end{figure*}

\subsection*{C. Sensitivity of the microwave quantum receiver}

The sensitivity of the microwave quantum receiver was evaluated by employing the experimental setup similar to the trEPR measurements. The key differences were the optical pump power and the condition of the microwave cavity optimized for facilitating the masing process and thus boosting the receiver's sensitivity. The optimized maser amplification was achieved when $Q_{\rm{L}}$ was increased from 1706 to 1857 and the cavity resonance was tuned to $\omega_{\rm{c}} = 2\pi \times 1.4492$ GHz. It is worth noting that higher $Q_{\rm{L}}$ should be avoided as it would result in maser oscillations that smear the microwave signals to be detected and thus are detrimental to the receiver's performance. According to the relationship $C \propto \sqrt{Q_{\rm{L}} / \omega_{\rm{c}}}$ validated in the trEPR experiment, the conversion factor $C$ was slightly modified to be 1.10 ± 0.03 mT/$\sqrt{\rm{W}}$ for analysing the sensitivity. 

By confirming the conversion factor, the sensitivity can be determined from the response of the receiver to calibrated $B_1$ fields along with the measured noise. A series of microwave signals with different powers were input into the receiver and the corresponding $B_1$ fields are shown in Fig. \ref{FIG.2}c. The response of the receiver $S$ (Fig. \ref{FIG.3}a) shows a nonlinear growth with the increased $B_1$  fields. Owing to the maser amplification, even for a nT-level weak $B_1$ field, it can be substantially enhanced and easily detected by the receiver in a single-shot measurement with a high signal-to-noise ratio (SNR) which is revealed by the small error bars in Fig. \ref{FIG.3}a. The feature of the maser amplification was characterized in terms of its gain as a function of frequency shown in the inset of Fig. \ref{FIG.3}a. Upon sweeping the frequency of the input microwave signals  $\omega_{\rm{in}}$, the cavity resonance was kept to be $\omega_{\rm{c}}=\omega_{\rm{in}}$ with the same coupling coefficient so that the $B_1$ fields of the input signals were ensured to be identical ($\sim 50$ nT) over the sweep range. The maximum gain was found to be 17 dB when $\omega_{\rm{in}}=2\pi\times1.4492$ GHz. The distribution of the gain at different frequencies follows the inhomogeneously broadened asymmetric EPR lineshape of pentacene triplet spin ensemble, which arises from the second-order hyperfine interactions at zero field \cite{doi:10.1126/sciadv.ade1613}. The negative gain, i.e. attenuation, observed on the wings of the distribution is due to the fact that the pentacene spin subensembles far from the EPR resonance are insufficient to produce enough gain to compensate the microwave energy trapped in the cavity and inevitable loss of the microwave circuitry (see the Supplemental Material for the gain and frequency analysis). As shown in Fig. \ref{FIG.3}a, the output signal is gradually saturated as the input microwave field increases, revealing that the receiver has the functionality of a limiter to protect the back-end from overload.

By differentiating the receiver's response to $B_1$ fields, the corresponding responsivity  $m_{\rm{s}} = \partial S / \partial B_1$ is obtained and shown in Fig. \ref{FIG.3}b. Relying on the definition of the sensitivity, $\eta = \frac{\sigma_{s}}{m_{\rm{s}} \sqrt{2\Delta{f}}}$ 
 \cite{doi:10.1126/sciadv.ade1613,PhysRevApplied.10.034044},
 in which $\sigma_{s}$ is the standard deviation of noise in single measurement and $\Delta{f}$ = 500 MHz is the equivalent noise bandwidth of our data acquisition. Thus the sensitivities under different microwave magnetic fields is determined as depicted in Fig. \ref{FIG.3}b; the optimal sensitivity is $1.18 \pm 0.16$ fT/$\sqrt{\rm{Hz}}$ . This is \textit{three} orders of magnitude lower than those achieved with NV sensors in diamond 
 \cite{Wang:2022klu,PhysRevApplied.19.054095}.
 We further verify the measured sensitivity by analysing the receiver's response to a weak microwave pulse with a frequency of $2\pi \times 1.4492$ GHz and a field strength $B_{\rm{test}}$ of 2.47 nT (Fig. \ref{FIG.3}c). We note that due to the dead time of the cavity, the input microwave signal was not only amplified but also prolonged. We obtain the SNR of the microwave power to be ${SNR}_{P}$=161.4, leading to the magnetic field ${SNR}_B$=12.7. 
Consequently, the experimental sensitivity (6.14 $\pm 0.17$ fT/$\sqrt{\rm{Hz}}$) of the receiver is obtained via $\eta = \frac{B_{\rm{test}}}{{
\rm{SNR}}\sqrt{2\Delta{f}}}$ \cite{Wang:2022klu}  which is on the same order of magnitude as the sensitivity predicted above based on the responsivity. Comparing with the reported microwave quantum receivers/magnetometers summarized in Supplementary Figure S2, the sensitivity of our device is superior to those of the NV-center based microwave magnetometers \cite{Wang:2022klu,2015High,Chen2023,PhysRevApplied.19.054095,PhysRevApplied.10.044039,Meinel2021,PhysRevApplied.12.044039} and thus can be considered as a promising room-temperature solid-state quantum device for microwave detection. On the other hand, we are also approaching so far the best sensitivity achieved by cold Rydberg atoms very recently \cite{2020Atomic,0Microwave,tu2023approaching} while revealing the robustness and capability of integration in contrast to the gaseous Rydberg systems.

\subsection*{D. Enhanced sensing via anharmonic quantum effects}

\begin{figure*}[htbp!]
\centering
\includegraphics[width=1\textwidth]{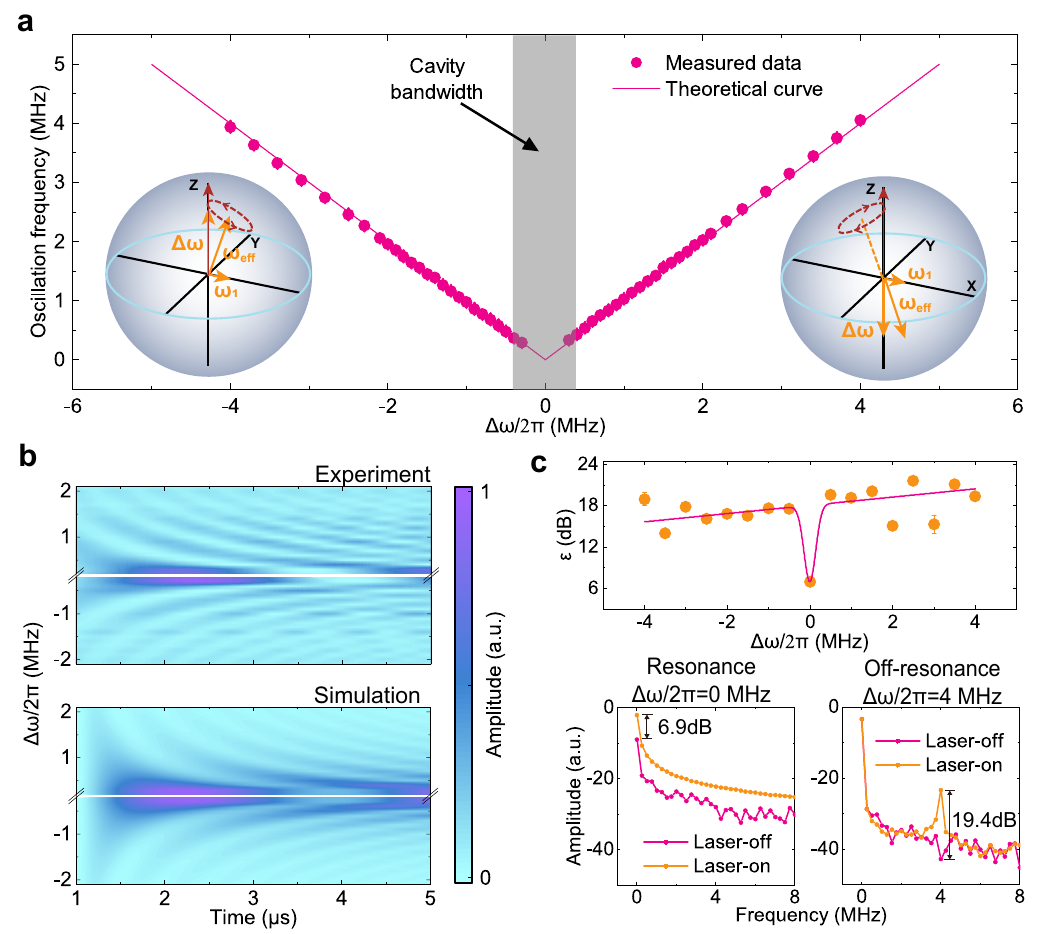}
\caption{\label{FIG.4}
\textbf{Proof-of-principle demonstrations of heterodyne detection of microwave fields.} (\textbf{a}) The beatings between the frequency of the cavity resonant mode $\omega_{\rm{c}}=2\pi\times1.4494$ GHz and the detuned input microwaves are mapped out with respect to the detuning frequency $\Delta\omega$. A linear relationship is displayed with a frequency range of ± 4 MHz which is far beyond the bandwidth of the cavity ($\sim$800 kHz). The error bars are contained within the data points. Inset: two Bloch spheres qualitatively describe the oscillations of polarized spin under negative (left) and positive (right) detunings. The yellow arrow represents microwaves and the red arrow represents the polarized spin, and the red dotted circle with the arrow represents the rotating direction of the spins. (\textbf{b}) The experimental and simulated maser response under varying the detuning frequency of the off-resonant microwave fields. The beatings with different detunings are mapped out as a function of time after applying laser pulses (upper) while the dynamics are simulated based on the driven Tavis–Cummings model(lower), see APPENDIX D. (\textbf{c}) The upper subplot represents the enhancement figure merit of heterodyne detection on microwave-field sensing. The bottom left and right subplots represent the FFT results for signals of laser-on (yellow) and -off (pink) under resonance (left) and off-resonance (right), respectively. The data points represent the mean ± standard deviation obtained from the five experiments.}
\end{figure*}

The aforementioned microwave detection with our receiver was implemented under a resonant condition, i.e. $\omega_{\rm{in}} = \omega_{\rm{c}}$, while for a practical receiver employed in most scenarios, the resonant condition is not always fulfilled. Therefore, it is necessary to evaluate the performance of the receiver when the frequencies of the input microwave signals mismatch with the cavity. We measured various microwaves applied to the receiver which were set within a range from 1.4454 GHz to 1.4534 GHz in steps of 100 kHz. Note that during our measurements, the cavity was always fixed to the pentacene triplet spin resonance despite the detunings between the cavity and input signals. As we adjusted the input frequency departure from the cavity resonance, beatings appeared with the output of the receiver. When the detuning increased, the frequency of the beatings grew linearly while the output amplitude was reduced as shown in Fig. \ref{FIG.4}a and b. The decreased amplitude arose from (i) the lowered maser gain at the off-resonant condition as demonstrated in the inset of Fig. \ref{FIG.3}a; (ii) the loss of the detected signal due to the detuning-induced reflection at the cavity input port which was characterized through the $S-$parameter measurement (see supplementary materials). The beat frequency is equal to the detuning $\Delta\omega = \omega_{\rm{c}} - \omega_{\rm{in}}$. The measured beat frequency agrees with the theoretical prediction (see APPENDIX D and Supplementary Materials for details) as shown in Fig. \ref{FIG.4}a which reveals that our quantum receiver is also capable of heterodyne measurements. More intriguingly, different from the reported quantum sensors, i.e. NV centers \cite{Wang:2022klu,PhysRevX.12.021061} and Rydberg atoms \cite{2020Atomic} that employ classical microwave sources for the local oscillators of heterodyne measurements, our hybrid quantum system by itself acts like the local oscillator for the frequency mixing. The underlying physics is the anharmonic  quantum interaction between an off-resonant microwave field and the pentacene triplet spin ensemble which can be described by the driven Tavis-Cummings model \cite{PhysRev.170.379,PhysRevLett.125.137701}  (see APPENDIX D for details), while the equality relation between frequencies of the beatings and detunings can be explained by solving Maxwell-Bloch equations \cite{hoff2012advanced} in the supplementary information.. 

In terms of the detection of the input frequency, since the duration of the input microwaves in our measurements were 5 $\mu$s limited by the lifetime of the masing process, the frequency of the beatings is unresolvable unless it is larger than 200 kHz whereas those signals with detuning smaller than 200 kHz can be observed with the prominent amplification near resonance. Despite the presence of the cavity functioning as a narrow-band filter, our detection range was not limited by the cavity bandwidth ($\approx$ 800 kHz). Strikingly, the valid detuning range exceeds the cavity bandwidth by approximately an order of magnitude, i.e. $\pm$ 4 MHz.  Though we only scan detuning up to $\pm$ 4 MHz, in principle, the beat frequency could reach the whole range  (about $\pm10$ MHz \cite{doi:10.1126/sciadv.ade1613}), where the zero-field magnetic resonance of pentacene triplet spin ensemble is measurable, i.e. sufficient number of spins presents in the range of the input frequency.

As the incident microwaves are measured by observing the output power of the maser-amplified signal, i.e. the maser burst's envelope which is measured directly by a logarithm detector, the modulated oscillations of the envelopes resulting from the anharmonic quantum effects circumvent the $1/f$ noise in contrast to the resonant detection and intuitively benefit the receiver's sensitivity. To verify this, we introduce a figure of merit, $\varepsilon$ to compare the on- and off-resonance detection enhancement offered by the receiver (see APPENDIX C). $\varepsilon$ was obtained from the ratio between the fast Fourier transform (FFT) amplitudes measured with and without the laser excitation, i.e.  the receiver was switched on and off.   
As shown in Fig. \ref{FIG.4}c, compared with the resonant detection, the heterodyne detection under the detuning conditions is improved and the maximum enhancement of $\varepsilon=19.4$ dB was achieved at $\Delta\omega/2\pi=4$ MHz whereas $\varepsilon$ obtained on resonance was 12.5 dB lower. In addition, we found that $\varepsilon$ measured at $\Delta\omega > 0$ is overall larger than that measured when $\Delta\omega < 0$. This was attributed to the asymmetric EPR lineshape of the pentacene triplet spin ensemble in which the spin subensembles are distributed preferentially in the regime $\Delta\omega > 0$ rather than $\Delta\omega < 0$ due to the second-order hyperfine interactions \cite{kohler1999magnetic}. Although the anharmonic-effect-assisted receiver possesses similar features to the heterodyne detectors, i.e. high sensitivity and the detected signals originate from the frequency beating, our approach effectively expands the detection bandwidth, especially lifts the restriction of the setup by the cavity whereas the traditional heterodyne technique sacrifices the detection bandwidth for the sensitivity. 

\section{III. Conclusion}
We have demonstrated a microwave quantum receiver based on the pentacene triplet spin ensemble coupled to a microwave cavity at ambient conditions. The detection scheme of the receiver is to measure microwave \textit{magnetic} fields in contrast to the traditional $\lambda/2$ dipole antennas and the Rydberg atomic receivers which are essentially electrometers. By exploiting the maser amplification in the solid-state hybrid quantum system at about 1.45 GHz, the receiver's sensitivity of $6.14 \pm 0.17$ $\rm{fT}/\sqrt{\rm{Hz}}$ is achieved. This figure surpasses the state-of-the-art microwave magnetometers based on solid-state spins of NV centers \cite{Wang:2022klu,PhysRevApplied.19.054095}. 
Moreover, the anharmonic quantum effects in the hybrid quantum system enable us to detect microwave signals spanning outside the cavity bandwidth by monitoring the heterodyne-like frequency beats with high SNRs. The counter-intuitive behavior successfully removes the restriction of the detection bandwidth set by the cavity and offers the enhanced detection capability under far-detuning conditions compared to the resonant scenarios. Unlike the heterodyne measurements demonstrated using the Rydberg atoms and NV centers that often require an external source to provide a local microwave field, our receiver operates without the need of such external microwave sources, which renders unique advantages, e.g. convenience and simplicity.

Our device possessing both functionalities of quantum mixers and amplifiers enriches the quantum toolbox for ultrahigh-sensitivity microwave detection and provides promising approaches for improving current microwave receiving systems in radars \cite{doi:10.1126/science.1160627}, wireless communications \cite{Koenig2013}, radio telescopes \cite{Pastor-Marazuela2021}, microwave circuitry imaging 
\cite{Anderson-microwave-circuit} and ESR/EPR \cite{poole1983electron,hoff2012advanced}. Our scheme, incorporating the amplification characteristic of masers into the sensing of microwave fields, can be extended to various mature solid-spin quantum systems, e.g. NV centers \cite{Breeze2018,jin2015proposal}, silicon vacancies in silicon carbide \cite{kraus2014room,fischer2018highly} and negatively charged boron vacancies in the van der Waals crystal hexagonal boron nitride \cite{gottscholl2020initialization,arroo2021perspective}, provided they can generate maser signals using appropriate polarization methods and suitable cavities. 

\section*{ACKNOWLEDGMENTS}
 We thank Ya Wang for valuable discussions. This study was supported by NSF of China (Grant No. 12374462, No. 12004037, No. 91859121, No. 12204040, No. 12321004), the National Key R\&D Program of China (Grant No. 2018YFA0306600), Beijing Institute of Technology Research Fund Program for Young Scholars and the China Postdoctoral Science Foundation (Grant No. YJ20210035, No. 2021M700439, No. 2023T160049), EPSRC New Horizons grant EP/V048430/1.

\section*{APPENDIX A: Experimental setup}
Configured as a cavity amplifier operated in reflection through a circulator, the setup of the microwave receiver is shown in supplementary materials. By utilizing a finite-element-method software, we designed the microwave cavity which was constracted from a hollow cylindrical single-crystal of STO containing a 0.1$\%$ pentacene-doped \textit{p}-terphenyl crystal (7.2 mm height, 4 mm width, 1.5 mm in average thick). The STO hollow cylinder was lifted upon a Rexolite support and housed within a cylindrical copper enclosure. A copper tuning piston on the top of the cavity was used to change the distance between the STO cylinder and the ceiling of the cavity to adjust the resonant frequency of the TE$_{01\delta}$ mode. The microwave resonator was directly coupled to a circulator using a small loop antenna as an undercoupled input/output port. By splitting the amplified microwave signals with a power splitter, we directly monitored the amplified microwave fields with a digital storage oscilloscope (Tektronix MSO64; sampling rate 6.25 GSa/s) whereas the power of the amplified microwave signals were measured by a logarithmic detector (AD8317; scale, 22 mV/dB) AC-coupled to the oscilloscope.

An optical parametric oscillator, OPO (Deyang Tech. Inc., BBOPO-Vis; pulse duration, 7 ns) pumped by its own internal Q-switched Nd:YAG laser was used to excite the Pc:Ptp crystal. The injected laser-pumping pulses as well as the test microwave fields generated from a MW source (SynthHDPro V2) were triggered by an ASG  (arbitrary sequence generator, CIQTEK, ASG8400).




\section*{APPENDIX B: On-resonance microwave magnetic field detection}
The microwave cavity was characterized by a microwave analyzer (Keysight, N9917A, see Supplementary materials). The measurement setups for the trEPR, maser gain and microwave magnetic field detection are the same, as described above. The microwave power used for trEPR measurement was set from -14.98 dBm to 6.67 dBm. The laser pulse used for the measurements of trEPR, maser gain and microwave magnetic field detection was monitored by a photodetector (Thorlabs, DET10A2).

\section*{APPENDIX C: Heterodyne detection}

We perform heterodyne detection with a detuning microwave fields. The power of input microwave generated by microwave source was -56.8 dBm and the frequency was varied in a range of 1.4454 GHz to 1.4534 GHz. To verify that heterodyne detection can circumvent 1/f noise, we introduced enhancement figure of merit and defined as
\begin{align}
\varepsilon = \rm{\frac{FFT(Laser-ON)}{FFT(Laser-OFF)}}\bigg|_{\Delta\omega}. 
\end{align}

\section*{APPENDIX D: Cavity electrodynamics of maser process}
In order to give a dynamical description of the maser process with detuning in Fig. \ref{FIG.4}, we start from the driven Tavis–Cummings Hamiltonian in the rotating frame 
 \cite{PhysRev.170.379,PhysRevLett.125.137701}. 
\begin{align}
     H = &\hbar\Delta_{\rm{c}}a^{\dagger} a + \frac{\hbar}{2}\Delta_{\rm{s}} \sum\limits_{j=1}^{N}\sigma_{j}^{z} +\hbar\sum\limits_{j=1}^{N}[g_{{j}}\sigma_{j}^{-}a^{\dagger} + g_{{j}}^{*}\sigma_{j}^{+}a] \notag\\
   &+ i\hbar [\eta a^{\dagger} - \eta^{*} a ], \label{H}
\end{align}
where $\Delta_{\rm{c}} =\omega_{\rm{c}}  - \omega_{\rm{in}}$ and $\Delta_{\rm{s}} =\omega_{\rm{s}} - \omega_{\rm{in}}$ are the detunings of the resonator frequency $\omega_{\rm{c}}$ and of the spin frequencies $\omega_{\rm{s}}$ from the input frequency $\omega_{\rm{in}}$ of the incoming microwave pulse with amplitude $\eta$. $a^{\dagger}$($a$) is the creation (annihilation) operator for the resonator mode coupling with $g_{j}$ to the $j$th spin that is presented with the Pauli operators $\sigma_{j}^{z}$, $\sigma_{j}^{+}$ and $\sigma_{j}^{-}$. 

The evolution of systems can be described according to quantum master equation, which can be written as 
\begin{equation}
\frac{\rm{d} \rho}{\rm{d}t}=-\frac{1}{i\hbar}[\rho,H]+\mathcal{L}[\rho]
\label{M}
\end{equation}
where $\mathcal{L}[\rho]$ stands for the Liouvillian \cite{2014Statistical, breeze2017room}
\begin{align}
   &\mathcal{L}[\rho]=\kappa_{\rm{c}}(a^{\dagger}\rho a-\frac{1}{2}(a^{\dagger}a\rho+\rho a^{\dagger}a)) + \gamma_{\parallel}\sum\limits_{j=1}^{N}(\sigma_{j}^{z}\rho\sigma_{j}^{z} - \rho) \notag\\
   &+\gamma_{\bot}\sum\limits_{j=1}^{N} (2\sigma_{j}^{-}\rho\sigma_{j}^{+} -\sigma_{j}^{+}\sigma_{j}^{-}\rho - \rho\sigma_{j}^{+}\sigma_{j}^{-}). \label{L} 
\end{align}
The first term in \eqref{L} describes the cavity losses with the decay rate $\kappa_{\rm{c}}$, the second and third term represent for the longitudinal and transversal decay rates $\gamma_{\parallel}$ and $\gamma_{\bot}$, respectively.

In the large spin ensembles limit, correlations between the cavity mode and spins can be neglected \cite{PhysRevLett.125.137701}, e.g. $\langle \sigma_{j}^{+} a\rangle \approx \langle \sigma_{j}^{+}\rangle\langle a\rangle$. According to the Lindblad master equation (2,3) and the approximate conditions above, we obtain a set of first-order differential equations for the expectation values of cavity mode field $\langle a\rangle$ and spin angular momentum $\langle \widetilde{S}_{-}\rangle$ and $\langle\widetilde{S}_{z}\rangle$:
\begin{align}
    &\frac{\rm{d}}{\rm{d}t}\langle a\rangle = -[\frac{1}{2}\kappa_{\rm{c}}+i\Delta_{\rm{c}}]\langle a\rangle -ig_{\rm{eff}}\langle\widetilde{S}_{-}\rangle  + \eta\\
    &\frac{\rm{d}}{\rm{d}t}\langle \widetilde{S}_{-}\rangle = -[\gamma_{\bot}+i\Delta_{\rm{s}}]\langle\widetilde{S}_{-}\rangle + ig_{\rm{eff}}\frac{1}{N}\langle \widetilde{S}_{z}\rangle \langle a \rangle\\
    &\frac{\rm{d}}{\rm{d}t}\langle\widetilde{S}_{z}\rangle = -\gamma_{\parallel}\langle{\widetilde{S}}_{z}\rangle + 2ig_{\rm{eff}}\frac{1}{N}(\langle\widetilde{S}_{-}\rangle\langle a^{\dagger}\rangle-\langle\widetilde{S}_{+}\rangle\langle a\rangle)
\end{align}
where $\widetilde{S}_{\pm}=\frac{1}{\sqrt{N}}\sum\limits_{j=1}^{N}\sigma_{j}^{\pm}$ and $\widetilde{S}_{z}=\frac{1}{{N}}\sum\limits_{j=1}^{N}\sigma_{j}^{z}$ are collective spin operators, and $g_{\rm{eff}}=g\sqrt{N}$ represents coupling strength which is derived under an assumption that spin-photon coupling of each spin is same and is a real number, i.e $g_j=g_j^{*}=g$ for any $j$. $\langle O\rangle$ is the expectation value of the operator $O$. Numerical solutions of the set of differential equations (4-6) are achieved by utilizing the Runge-Kutta method. To fit the experimental results measured by a logarithmic detector in Fig. \ref{FIG.4}a, the simulated results shown in Fig. \ref{FIG.4}b was calculated by $\lg\langle a^{\dagger}a\rangle$ and normalized by its maximum and minimum value. In the simulation, an approximation $\langle a^{\dagger}a\rangle \approx \langle a^{\dagger}\rangle\langle a\rangle = |\langle a\rangle|^2$ is adopted, where $\langle a\rangle$ obtained from the solutions of equations (4-6).

\nocite{*}

\bibliography{apssamp}

\end{document}